\documentclass[%
reprint,
groupedaddress,
amsmath,amssymb,
aps,
prd,
]{revtex4-2}

\usepackage{graphicx}
\usepackage{dcolumn}
\usepackage{bm}

\usepackage{amsfonts}
\usepackage{slashed}
\begin{document}
\title{Erratum: Exact solutions to the fermion propagator Schwinger-Dyson equation in Minkowski space with on-shell renormalization for quenched QED \newline [Phys. Rev. D 96, 036021 (2017)]}
\author{Shaoyang Jia}
\email{sjia@iastate.edu}
\affiliation{Department of Physics \& Astronomy, Iowa State University, Ames, Iowa 50010, USA}
\affiliation{Physics Division, Argonne National Laboratory, 9700 South Cass Avenue, Lemont, Illinois 60439, USA}
\author{Michael R. Pennington}
\affiliation{School of Physics \& Astronomy, Glasgow University, Glasgow G12 8SU, UK}
\maketitle
\section{SDE for the fermion propagator spectral functions\label{ss:solution_xi0}}
In finding the solution of the fermion propagator with the correct analytic structure from its Schwinger--Dyson equation (SDE) in four dimension, we applied a modeling with a purely \lq\lq longitudinal'' vertex~\cite{Delbourgo:1977jc,PhysRev.130.1287,PhysRev.135.B1398,PhysRev.135.B1428} in quenched quantum electrodynamics (QED). The advantage with this modeling is that solutions can be found explicitly in series form. However such an Ansatz is known to violate gauge covariance~\cite{Delbourgo:1980vc,Delbourgo:1979eu,Jia:2016udu,Pennington:2016vxv,Jia:2016wyu}. Consequently the resulting SDEs for the fermion propagator can only be renormalized on-shell in one particular gauge. In previous work we incorrectly took this to be the Landau gauge~\cite{Jia:2017niz}. This erratum corrects such a mistake, now working in the Yennie gauge. Because solutions in different gauges will not be consistent with each other, the Euclidean space fermion propagator in the Landau gauge is obtained using the Landau--Khalatnikov--Fradkin transform (LKFT)~\cite{Jia:2016udu,Jia:2016wyu}.

We improve the on-shell renormalization conditions in Subsection~IV.B of Ref.~\cite{Jia:2017niz} by introducing $r_0$ as the residue of the free-particle pole, the default choice of which is unity. Equation~(37) of Ref.~\cite{Jia:2017niz} then becomes
\begin{subequations}\label{eq:OnShellCondj}
\begin{align}
	& S_1(p^2) = \dfrac{r_0}{p^2-m^2}+P_1(p^2)\label{eq:OnShellCond1}\\
	& S_2(p^2) = \dfrac{m\,r_0}{p^2-m^2}+P_2(p^2),\label{eq:OnShellCond2}
\end{align}
\end{subequations}
where $P_j(p^2)$ is less singular than the free-particle propagator in the vicinity of $m^2$. Such an on-shell renormalization condition indicates
\begin{subequations}\label{eq:rhoj_delta_theta}
	\begin{align}
		& \rho_1(s) = r_0\,\delta(s-m^2)+r_1(s),\label{eq:rho1_delta_theta}\\
		& \rho_2(s) = m\, r_0\,\delta(s-m^2)+r_2(s),\label{eq:rho2_delta_theta}
	\end{align}
\end{subequations}
where $r_{j}(s)$ are expected to be regular functions. Equation~(39) of Ref.~\cite{Jia:2017niz} is improved into
\begin{subequations}\label{eq:def_bbSj}
\begin{align}
	& \overline{\sigma}_1(p^2) = -\dfrac{\lambda_1\alpha}{4\pi}\,\dfrac{m^2}{p^2-m^2}+q_1(p^2),\label{eq:def_bbS1}\\
	& \overline{\sigma}_2(p^2) = -\dfrac{\lambda_2\alpha}{4\pi}\,\dfrac{m}{p^2-m^2}+q_2(p^2).\label{eq:def_bbS2}
\end{align}
\end{subequations}
Following steps in App.~\ref{ss:simplification_SDE_fermion} and the separation of the free-propagator terms, Eq.~(40) of Ref.~\cite{Jia:2017niz} becomes
\begin{subequations}\label{eq:SDE_rj_OnShell}
\begin{align}
	& p^2P_1(p^2)+q_1(p^2) + (\lambda_2-\lambda_1)\dfrac{\alpha}{4\pi}\dfrac{m^2}{p^2-m^2} \nonumber\\
	& = \left( 1 - \dfrac{\lambda_2\alpha}{4\pi r_0} \right) mP_2(p^2) + (\lambda_2-\lambda_1) \dfrac{\alpha}{4\pi} \lim\limits_{\mu^2\rightarrow m^2} \dfrac{m^2}{\mu^2-m^2} \nonumber\\
	& + \left( 2-\dfrac{\lambda_2\alpha}{4\pi r_0} \right) [ m^2P_1(m^2)-mP_2(m^2) ] + q_1(m^2) \nonumber\\
	& - m\, q_2(m^2) \label{eq:SDE_r1_OnShell}\\
	& P_2(p^2)+q_2(p^2) = \left( 1 - \dfrac{\lambda_2\alpha}{4\pi r_0} \right) m P_1(p^2),\label{eq:SDE_r2_OnShell}
\end{align}
\end{subequations}
when the strength of the free-particle pole is adjustable. Applying Eqs.~\eqref{eq:rhoj_delta_theta} and~\eqref{eq:def_bbSj}, Eq.~(42) of Ref.~\cite{Jia:2017niz} becomes
\begin{subequations}
\begin{align}
	& q_1(p^2) = \overline{\sigma}_1(p^2) + \dfrac{ \lambda_1\alpha }{ 4\pi r_0 } \dfrac{m^2}{p^2-m^2} = - \dfrac{ 3\alpha r_0 }{ 4\pi } \dfrac{m^2}{p^2} \ln\dfrac{m^2}{m^2-p^2} \nonumber\\
	& + \dfrac{ \alpha\xi r_0 }{ 4\pi } \left( 1 + \ln\dfrac{m^2}{m^2-p^2} \right) - \dfrac{ 3\alpha }{ 4\pi } \int_{m^2}^{+\infty}ds \dfrac{ s\overline{K}(p^2,s) }{ p^2-s } r_1(s) \nonumber\\
	& + \dfrac{\alpha\xi}{4\pi} \int_{m^2}^{+\infty}ds \left( 1 + \ln\dfrac{m^2}{s-p^2} \right) r_1(s),\label{eq:def_q_1} \\
	& q_2(p^2) = \overline{\sigma}_2(p^2) + \dfrac{\lambda_2\alpha}{4\pi r_0}\dfrac{m}{p^2-m^2} = - \dfrac{3\alpha r_0}{4\pi}\dfrac{m}{p^2} \ln\dfrac{m^2}{m^2-p^2} \nonumber\\
	& - \dfrac{ \alpha\xi r_0 }{4\pi} \dfrac{m}{p^2} \Big( 1 - \dfrac{m^2}{p^2}\ln\dfrac{m^2}{m^2-p^2} \Big) - \dfrac{3\alpha}{4\pi} \int_{m^2}^{+\infty}ds \dfrac{\overline{K}(p^2,s)}{p^2-s} \nonumber\\
	& \times r_2(s) - \dfrac{\alpha\xi}{4\pi} \int_{m^2}^{+\infty}ds \dfrac{1}{p^2} \Big( 1 - \dfrac{s}{p^2}\ln\dfrac{s}{s-p^2} \Big) r_2(s).
\end{align}
\end{subequations}
Notice that the $p^2$ dependences of $q_1(p^2)$ in Eq.~\eqref{eq:def_q_1} cannot be expressed in the spectral representation without subtraction. See App.~\ref{ss:sub_Diarc_scalar} for the subtraction of Eq.~\eqref{eq:def_q_1} at $p^2=0$. 

Next in order to deduce the equations for $r_j(s)$ of Eq.~\eqref{eq:rhoj_delta_theta}, we reproduce the $p^2$ dependences in Eq.~\eqref{eq:SDE_rj_OnShell} using the spectral representation. Because $q_2(p^2)$ and the subtracted loop integral $\tilde{q}_1(p^2)$ defined in Eq.~\eqref{eq:tilde_q1} are corrections to the fermion propagator, equations for $r_j(s)$ can also be deduced by taking imaginary parts along the branch cuts of the logarithmic functions directly. As a correction to Eq.~(43) of Ref.~\cite{Jia:2017niz} due to an incorrect inhomogeneous term, both methods lead to the following integral equations for $r_j(s)$:
\begin{subequations}\label{eq:SDE_R_OnShell}
\begin{align}
	& s^2\, r_1(s;\xi) - ms\, r_2(s;\xi) + a \left[ m^2 + \int_{m^2}^{s}ds'\, s' r_1(s';\xi) \right]\nonumber\\
	& = bs \left[ 1 + \int_{m^2}^{s}ds'\, r_1(s';\xi) \right] \\
	& - m s^2\, r_1(s;\xi) + s^2\, r_2(s;\xi) + as \left[ m + \int_{m^2}^{s}ds'\, r_2(s';\xi) \right] \nonumber\\
	& = b \left[ m^3 + \int_{m^2}^{s}ds's'\, r_2(s';\xi) \right],
\end{align}
\end{subequations}
where we have defined 
\begin{equation}
	a \equiv \dfrac{3\alpha}{4\pi} \dfrac{r_0}{ 1 - \alpha/\pi }, \quad\mathrm{and}\quad b \equiv \dfrac{ \alpha\xi }{ 4\pi } \dfrac{r_0}{ 1 - \alpha/\pi }.
\end{equation}
Because $p^2=0$ is chosen as the subtraction point for the Dirac scalar component of the SDE, we also need to supplement Eq.~\eqref{eq:SDE_R_OnShell} by Eq.~\eqref{eq:SDE_r1_OnShell_p20_rdd} to recover Eq.~\eqref{eq:SDE_r2_OnShell}.

Equation~\eqref{eq:SDE_R_OnShell} is the updated SDE for fermion spectral functions $\rho_{j}(s)$ with a loop-renormalizable modification to the Gauge Technique in the quenched approximation. Because spectral variables $s$ and $s'$ are separable, these integral equations can be converted into differential equations by taking derivatives with respect to $s$. In the Yennie gauge, two derivatives are required. In the next section, we will discuss how to find solutions for Eq.~\eqref{eq:SDE_R_OnShell} in this gauge.
\section{Solutions in the Yennie gauge\label{ss:Yennie}}
\subsection{Conversion into differential equations}
By taking derivatives with respect to the spectral variable $s$, Eq.~\eqref{eq:SDE_R_OnShell} can be converted into first order differential equations in the Landau gauge where $\xi=0$, in this case there should be at most two linearly independent solutions. Whereas in any other covariant gauges, Eq.~\eqref{eq:SDE_R_OnShell} can be converted into second order differential equations, where there are four linearly independent solutions. The initial conditions for the differential equations are recovered from the limiting behavior of the original integral equation. We will demonstrate explicitly that in the Landau gauge, there is no solution with finite initial conditions for the spectral functions $r_j(s;0)$. Therefore the SDE for the fermion propagator using the Gauge Technique Ansatz, with a loop-renormalizable modification, cannot be renormalized on-shell in the Landau gauge with non-vanishing coupling constant. Importantly this corrects the algebra in Eq.~(A12) of Ref.~\cite{Jia:2017niz}. Unfortunately the inhomogeneous term was mistakenly added to Eq.~(A15a) of Ref.~\cite{Jia:2017niz}, leading to a wrong result. We should have known from Ref.~\cite{Eides:2001dw} that only in the Yennie gauge was on-shell renormalization consistent. Here we give the correct results. In fact, we will see explicitly that within this Ansatz, there is uniquely one choice of the gauge fixing parameter, under which the on-shell renormalization conditions can be consistently implemented.

In order to find out which gauge is allowed by Eq.~\eqref{eq:SDE_R_OnShell}, let us start by analyzing the behavior of $r_j(s;\xi)$ in the limit of $s\rightarrow m^2$. For notational convenience, we define the following dimensionless functions
\begin{equation}
	f_1(y) = s\,r_1(s) \big\vert_{s=m^2 y} ~\mathrm{and}~ f_2(y) = s r_2(s)/m^2 \big\vert_{s=m^2 y}
\end{equation}
of the variable $y=s/m^2$. Equation~\eqref{eq:SDE_R_OnShell} then is equivalently written as
\begin{subequations}\label{eq:itg_f1_f2}
\begin{align}
	& yf_1(y) - f_2(y) + a \left[ 1 + \int_{1}^{y}dy'\,f_1(y') \right] \nonumber\\
	& = by \left[ 1 + \int_{1}^{y}dy'\, \dfrac{f_1(y')}{y'} \right], \\[1mm]
	& - y f_1(y) + yf_2(y) + ay \left[ 1 + \int_{1}^{y}dy'\, \dfrac{f_2(y')}{y'} \right] \nonumber\\
	& = b \left[ 1 + \int_{1}^{y}dy'\, f_2(y') \right].
\end{align}
\end{subequations}
In the limit of $y\rightarrow 1$, we obtain
\begin{equation}
\begin{cases}
	f_1(1)-f_2(1)+a=b \\
	-f_1(1)+f_2(1)+a=b
\end{cases},\label{eq:f_12_1}
\end{equation}
which requires
\begin{equation}\label{eq:cond_f_12_1}
	a=b,\quad \mathrm{and}\quad f_1(1)=f_2(1).
\end{equation}
Consequently if $a\neq b$ there is no consistent solution to Eq.~\eqref{eq:f_12_1}. Since $a=b$ only happens in the Yennie gauge when $\xi=3$, away from this particular gauge there is no consistent solution to Eq.~\eqref{eq:SDE_R_OnShell} without violating the on-shell renormalization condition. Therefore the Gauge Technique only allows on-shell renormalization for the fermion propagator in the Yennie gauge. When $\alpha=0$, the theory is trivial. While when $\alpha=\pi$, there is no consistent solution even in the Yennie gauge. Therefore we only consider the numerical solutions when $0<\alpha<\pi$. 

To see that in the Yennie gauge the limiting behaviors of $r_{j}(s)$ are indeed well-defined, let us take derivatives with respect to $y$ on Eq.~\eqref{eq:itg_f1_f2} at the $y\rightarrow 1$ limit. Essentially we are looking for the Taylor series solutions expanded about the point $y=1$. Specifically taking the first derivative on Eq.~\eqref{eq:itg_f1_f2} produces
\begin{subequations}\label{eq:Df_12_1_ori}
\begin{align}
	& yf_1'(y)+f_1(y)-f_2'(y)+af_1(y) \nonumber\\
	& = b \left[1+\int_{1}^{y}dy'\dfrac{f_1(y')}{y'} \right]+bf_1(y), \\[1mm]
	& -y f_1'(y)-f_1(y)+yf_2'(y)+f_2(y) \nonumber\\
	& + a \left[1+\int_{1}^{y}dy'\dfrac{f_2(y')}{y'} \right]+af_2(y)=bf_2(y).
\end{align}
\end{subequations}
The subsequent application of Eq.~\eqref{eq:cond_f_12_1} simplifies Eq.~\eqref{eq:Df_12_1_ori} into
\begin{subequations}\label{eq:Df_12_1}
\begin{align}
	& y f_1'(y) + f_1(y) - f_2'(y) = b \left[ 1 + \int_{1}^{y}dy'\dfrac{f_1(y')}{y'} \right], \\
	& - y f_1'(y) - f_1(y) + y f_2'(y) + f_2(y) \nonumber\\
	& + a \left[ 1 + \int_{1}^{y}dy'\dfrac{f_2(y')}{y'} \right] = 0.
\end{align}
\end{subequations}
In the limit of $y\rightarrow 1$, Eq.~\eqref{eq:Df_12_1} simplifies into 
\begin{equation}
\begin{cases}
f_1(1)=0\\
f_1'(1)-f_2'(1)=a
\end{cases}.\label{eq:cond_Df_12_1}
\end{equation}
Taking the derivative with respect to $y$ on Eq.~\eqref{eq:Df_12_1} produces 
\begin{equation}
\begin{cases}
yf_1''(y)+2f_1'(y)-f_2''(y)=bf_1(y)/y\\
-yf_1''(y)-2f_1'(y)+yf_2''(y)+2f_2'(y)+af_2(y)/y=0
\end{cases}.\label{eq:DDf_12}
\end{equation}
Again in the limit of $y\rightarrow 1$, we obtain
\begin{equation}
\begin{cases}
f_1'(1)=a\\
f_1''(1)-f_2''(a)=-2a
\end{cases}.\label{eq:cond_DDf_12_1}
\end{equation}

In summary, we have obtained the following boundary conditions for $f_j(y)$: 
\begin{equation}
\begin{cases}
f_1(1)=0,\quad f_2(1)=0,\\
f_1'(1)=a, \quad f_2'(1)=0,\\
f_2''(1)=f_1''(1)+2a.
\end{cases}\label{eq:initial_cond_Yennie}
\end{equation}
We only need the first four conditions as the initial values to find out the unique pair of $f_j(y)$ from their differential equations.
\subsection{The series solution in the Yennie gauge}
\subsubsection{The series expansion with $x=s/m^2-1$}
Having deduced the equations for $r_{j}(s;\xi=3)$ given by Eq.~\eqref{eq:DDf_12} with initial conditions Eq.~\eqref{eq:initial_cond_Yennie}, we could derive the recurrence relations for the Taylor series solutions. For convenience, we shift the starting point of the functions to the origin using $y=x+1$. We then write down the series expansions of $f_j$ on the variable $x$. Explicitly we have 
\begin{equation}
f_1(x)=\sum_{n=1}^{+\infty}a_nx^n \quad\mathrm{and}\quad 
f_2(x)=\sum_{n=2}^{+\infty}b_nx^n.
\end{equation}
We then substitute these expansions into Eq.~\eqref{eq:DDf_12} written equivalently as
\begin{subequations}
\begin{align}
& (x+1)^2\dfrac{d^2}{dx^2}f_1+2(x+1)\dfrac{d}{dx}f_1-(x+1)\dfrac{d^2}{dx^2}f_2=af_1\\
& x(x+1)\dfrac{d^2}{dx^2}f_2+2(x+1)\dfrac{d}{dx}f_2+af_2=af_1,
\end{align}
\end{subequations} 
and obtain the following recurrence relations:
\begin{subequations}
\begin{align}
	& a_1=a, \quad b_2-a_2=a, \quad b_2=a^2/6, \\
	& [n(n+1)-a]a_n+2(n+1)^2a_{n+1}+(n+1)(n+2)a_{n+2} \nonumber\\
	& = n(n+1)b_{n+1}-(n+1)(n+2)b_{n+2} \,\,\, \mathrm{for}~n\geq 1,\\
	& [(n+1)(n+2)+a]b_{n+1}+2(n+2)^2b_{n+2}=aa_{n+1}\nonumber\\
	& \quad \mathrm{for}~n\geq 1.
\end{align}
\end{subequations}
Although formally these series solutions for $f_j(x)$ have been found, in practice they only converge for small $x$, and so are not appropriate to account for the asymptotic behaviors of the spectral functions.
\subsubsection{The series expansion with $z=m^2/s$}
To understand the asymptotic behavior of the two spectral functions, let us consider another variable transform $z=m^2/s$. Specifically, we define 
\begin{equation}
g_1(z)=r_1(s)/m^2\big\vert_{s=m^2/z}, \quad g_2(z)=r_2(s)/m\big\vert_{s=m^2/z}.
\end{equation}
Equation~\eqref{eq:DDf_12} then becomes
\begin{subequations}
\begin{align}
& z^2\dfrac{d^2}{dz^2}g_1(z)-2z\dfrac{d}{dz}g_1(z)-z^3\dfrac{d^2}{dz^2}g_2(z) \nonumber\\
&\hspace{4cm}+(2-a)g_1(z)=0, \label{eq:Dg_12a} \\
& -z^2\dfrac{d^2}{dz^2}g_1(z)+2z\dfrac{d}{dz}g_1(z)+z^2\dfrac{d^2}{dz^2}g_2(z)\nonumber\\
&\hspace{1cm}-2z\dfrac{d}{dz}g_2(z)-2g_1(z)+(2+a)g_2(z)=0.\label{eq:Dg_12b}
\end{align}
\end{subequations}
Summing Eqs.~\eqref{eq:Dg_12a} and~\eqref{eq:Dg_12b} produces
\begin{equation}
ag_1(z)=(2+a)g_2(z)-2z\dfrac{d}{dz}g_2(z)+z^2(1-z)\dfrac{d^2}{dz^2}g_2(z).\label{eq:Dg_120}
\end{equation}
Substituting Eq.~\eqref{eq:Dg_120} into Eq.~\eqref{eq:Dg_12a}  gives
\begin{align}
& z^4(1-z)\dfrac{d^4}{dz^4}g_2(z)-4z^4\dfrac{d^3}{dz^3}g_2(z)+2z^2(1-z)\dfrac{d^2}{dz^2}g_2(z)\nonumber\\
& -4z\dfrac{d}{dz}g_2(z)+(4-a^2)g_2(z)=0,\label{eq:D^4g_2}
\end{align}
as a fourth-order ODE for $g_2(z)$. After realizing that $z=0$ is a regular singular point of Eq.~\eqref{eq:D^4g_2}, we write the following series solution:
\begin{equation}
g_2(z)=\sum_{n=0}^{+\infty}c_n^{(i)}z^{n+\gamma_i},\label{eq:series_g2}
\end{equation}
where $\gamma_i$ with $i\in \{1,2,3,4\}$ are the roots of
\begin{equation}
\gamma(\gamma-1)(\gamma-2)(\gamma-3)+2\gamma(\gamma-1)-4\gamma+(4-2a^2)=0.
\end{equation}
Explicitly, we have
\begin{align}
& \gamma_1=(3-\sqrt{1+4a})/2,\quad \gamma_2=(3+\sqrt{1+4a})/2,\nonumber\\
& \gamma_3=(3-\sqrt{1-4a})/2,\quad \gamma_4=(3+\sqrt{1-4a})/2.\label{eq:def_gamma_i}
\end{align}
There exists a simple recurrence relation for $c_n$:
\begin{equation}
\dfrac{c_{n+1}}{c_n}=\dfrac{(\gamma+n)^2(\gamma+n-1)^2}{(\gamma+n)^2(\gamma+n-1)^2-b^2}.
\end{equation}
Once $g_2(z)$ is known, $g_1(z)$ is given by Eq.~\eqref{eq:Dg_120}. 

Although we cannot easily match $c_n^{(i)}$ to the initial conditions given by Eq.~\eqref{eq:initial_cond_Yennie}, the leading term in Eq.~\eqref{eq:series_g2} specifies the asymptotic behavior of the $r_2(s)$ as
\begin{equation}
\lim\limits_{s\rightarrow +\infty}r_2(s)=c_0^{(i)}m^{2\gamma_i-1} s^{-\gamma_i}.\label{eq:asym_r2}
\end{equation}
Meanwhile, from Eq.~\eqref{eq:Dg_120} we obtain the asymptotic behavior for $r_1(s)$ as 
\begin{align}
& \lim\limits_{s\rightarrow+\infty} r_1(s) = c_0^{(i)} m^{2\gamma_i-2} s^{-\gamma_i} \nonumber\\
& \times \bigg\{ \dfrac{1}{a} \left(\gamma_i-\dfrac{3-\sqrt{1-4a}}{2}\right) \left(\gamma_i-\dfrac{3+\sqrt{1-4a}}{2}\right) \nonumber\\
& + \dfrac{4\gamma_i(1-\gamma_i)}{ \left[ \gamma_i - ( 1-\sqrt{1+4a} ) / 2 \right] \left[ \gamma_i - ( 1+\sqrt{1+4a})/2 \right] }\dfrac{m^2}{s}\nonumber\\
& + \mathcal{O}\left( m^2 / s \right)^2 \bigg\}.\label{eq:asym_r1}
\end{align}
We will confirm these asymptotic relations by fitting numeric solutions of $r_j(s)$.
\begin{figure*} 
\hspace*{-2.5cm}
\includegraphics[width=1.2\linewidth]{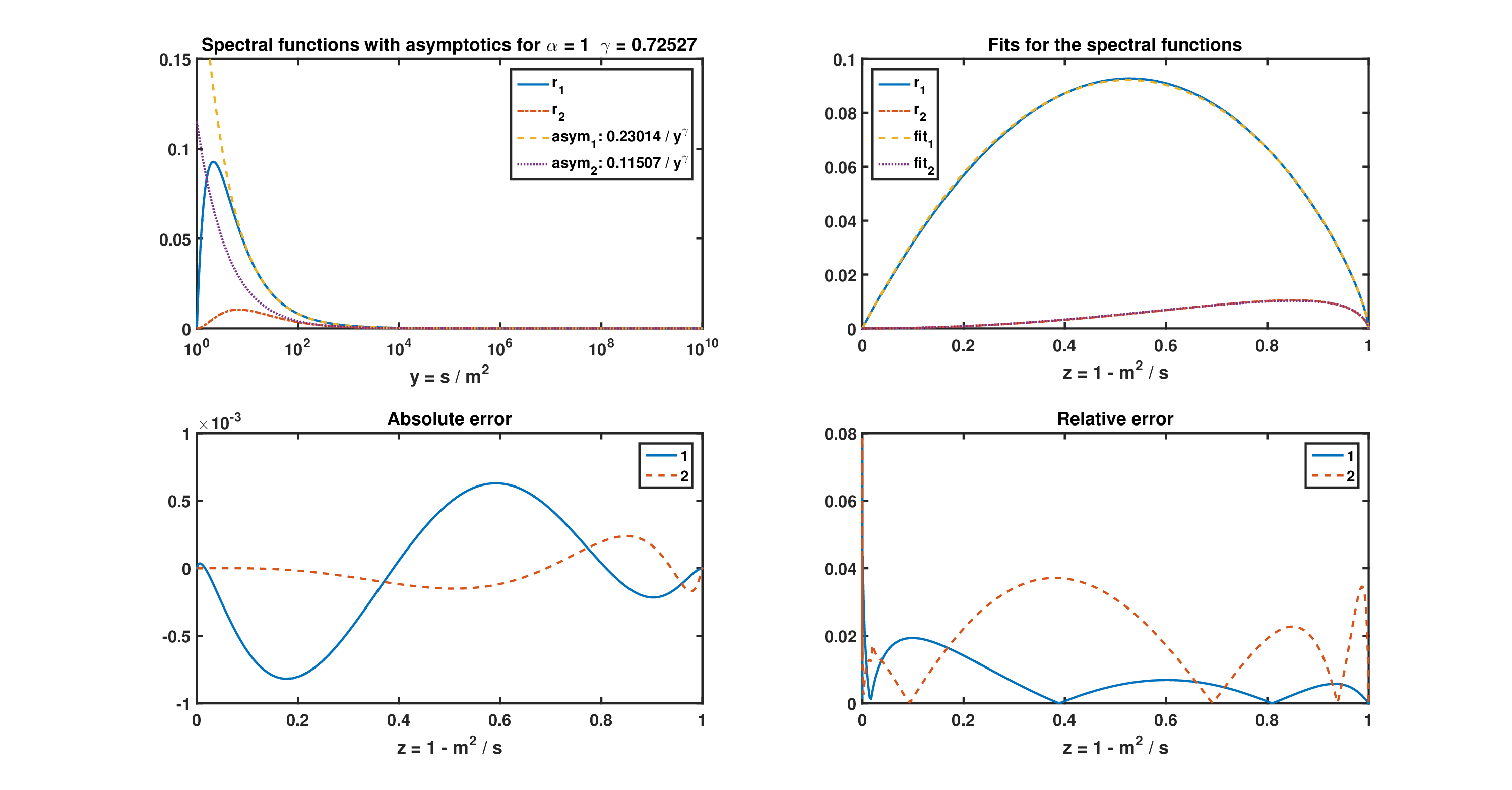}
\caption{The numerical solutions for the spectral functions $r_1$ and $r_2$ together with their Pad\'{e} approximations and errors of this parameterization with respect to numerical solutions. In the upper-left plot, the blue solid line stands for the numerical solution of $r_1(s)$. The red dot-dash line represents the numerical solution of $r_2(s)$. The yellow dash line and the purple dot line are the asymptotic power behaviors for $r_1(s)$ and $r_2(s)$, respectively. In the upper-top plot, the functions $r_j(s)$ and their parameterizations by Eq.~\eqref{eq:r_j_Pade} are illustrated. The blue solid line and the red dot-dash line are the numerical solutions for $r_1(s)$ and $r_2(s)$ as functions of $z=1-m^2/s$. The yellow dash line and the purple dot line are the Pad\'{e} approximations for $r_1(s)$ and $r_2(s)$ respectively. The Pad\'{e} approximations coincide closely to the numerical solutions. The lower two plots are the absolute and the relative errors of the Pad\'{e} approximations with respect to the numerical solutions of $r_j(s)$. The left plot represents the absolute errors and the right plot represents the relative errors. In both of the lower plots, the blue solid lines represent errors for $r_1(s)$, while the red dash lines are for $r_2(s)$.}
\label{fig:r1r2_n10}
\end{figure*}
\subsection{Numerical solutions and their parameterizations}
For computational convenience, let us define functions $u_1(y)$ and $u_2(y)$ as
\begin{align}
& u_1(y)=yf_1(y)-f_2(y)\\
& u_2(y)=y[f_1(y)-f_2(y)].
\end{align}
We then have $f_1(y)=[u_1(y)-u_2(y)/y]/(y-1)$, and $f_2(y)=[u_1(y)-u_2(y)]/(y-1)$. Next, Eq.~\eqref{eq:DDf_12} simplifies into 
\begin{subequations}\label{eq:D^2_u12_y}
\begin{align}
& \dfrac{d^2}{dy^2}u_1(y)=\dfrac{a}{(y-1)y}\left[u_1(y)-\dfrac{u_2(y)}{y} \right]\\
& \dfrac{d^2}{dy^2}u_2(y)=\dfrac{a}{(y-1)y}\left[u_1(y)-u_2(y) \right].
\end{align}
\end{subequations}
The initial conditions corresponding to Eq.~\eqref{eq:initial_cond_Yennie} are now
\begin{equation}
\begin{cases}
u_1(0)=0,\quad u_1'(0)=a,\\
u_2(0)=0,\quad u_2'(0)=a.
\end{cases}\label{eq:ini_u12}
\end{equation}
Note that derivative operators in Eq.~\eqref{eq:D^2_u12_y} are simpler than those in Eq.~\eqref{eq:DDf_12}. Consequently Eq.~\eqref{eq:D^2_u12_y} is more suitable to be solved numerically. 

\begin{figure*}
\hspace*{-2cm}
\includegraphics[width=1.2\linewidth]{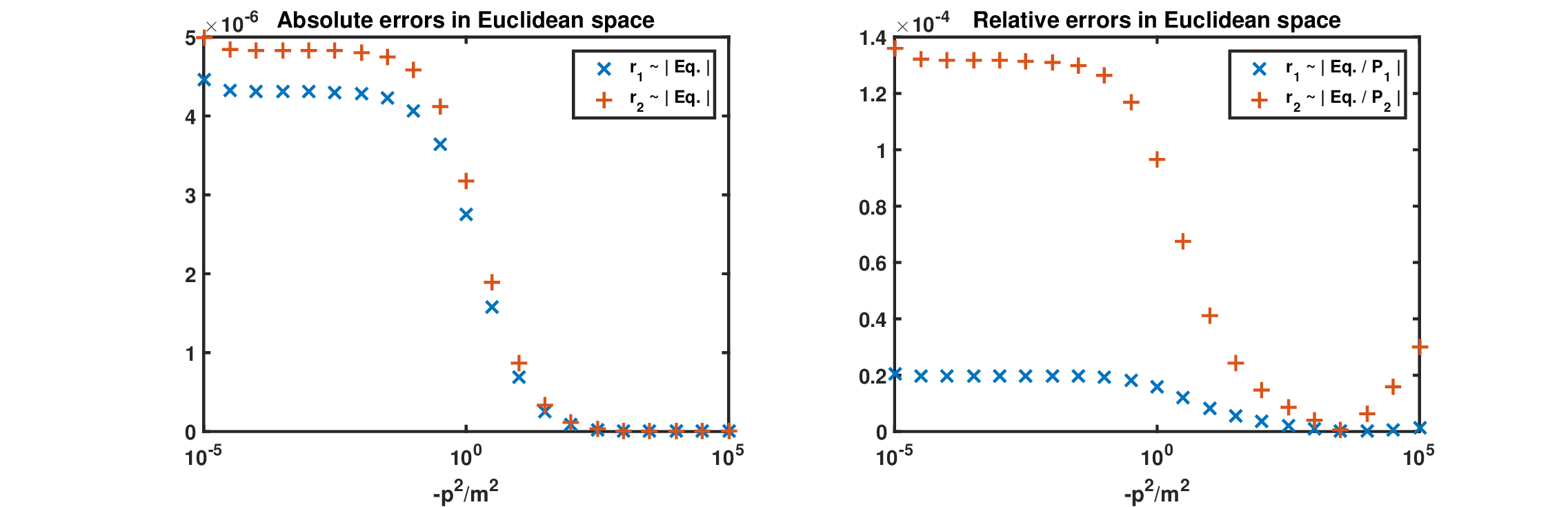}
\caption{Errors of the SDEs in the Euclidean space with the numerical solution at $\alpha=1$. Absolute errors are given in the left plot, while relative errors are shown on the right plot. Here the reference for the relative errors are chosen to be $P_j(p^2)$. In both plots, the blue crosses stand for errors from Eq.~\eqref{eq:Subt_Cond}, while the red pluses stand for errors from Eq.~\eqref{eq:SDE_r2_OnShell}.}
\label{fig:Euc_Err}
\end{figure*}

As an example, we present the numerical solution of the spectral functions $r_j(s)$ when $\alpha=1$ and $r_0=1$, given by the upper-left plot of Fig.~\ref{fig:r1r2_n10}. Specifically, we verify that the asymptotic behaviors match the power law corresponding to $s^{-\gamma_1}$ with $\gamma_1$ given by Eq.~\eqref{eq:def_gamma_i}. We also found out the proportionality constant $c_0^{(1)}=0.11507$ in Eqs.~\eqref{eq:asym_r1} and~\eqref{eq:asym_r2} by fitting the numerical asymptotics. We then substitute the numerical solutions into Eqs.~\eqref{eq:Subt_Cond} and~\eqref{eq:SDE_r2_OnShell} to test the numerical errors in Euclidean space. These errors are illustrated in Fig.~\ref{fig:Euc_Err}, within which the relative errors are calculated with respect to $P_j(p^2)$. 

Notice that the differential equations for $u_j(y)$ are homogeneous. While the initial conditions are proportional to $r_0$ with fixed $\alpha$. Other choices of $r_0$ simply scales the $r_0=1$ solutions. Additionally, we have verified numerically that the choices of $r_0$ is not constrained by the subtraction condition in Eq.~\eqref{eq:SDE_r1_OnShell_p20}. 
\begin{figure}
\vspace*{-1.15cm}
\includegraphics[width=1.0\linewidth]{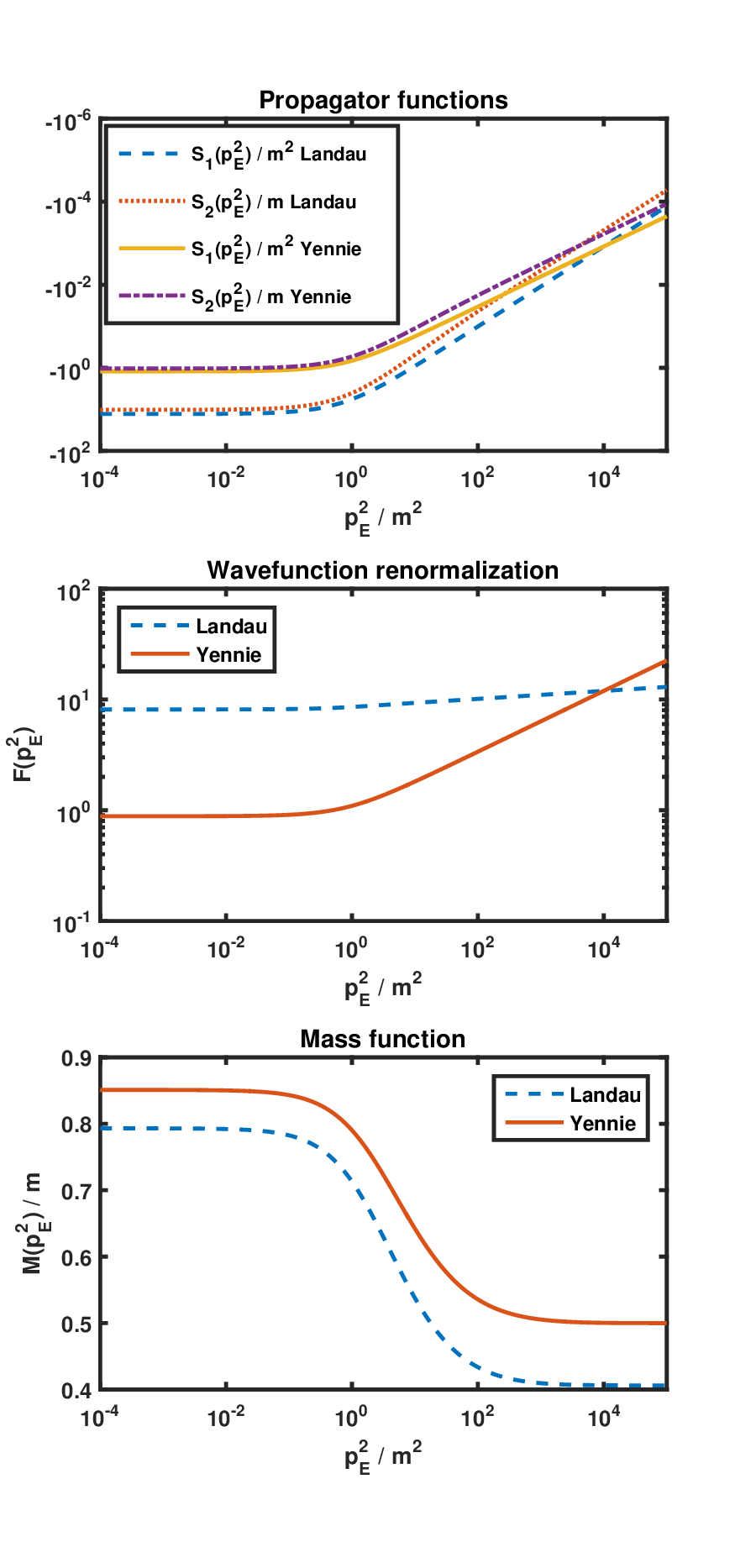}
\vspace*{-1.5cm}
\caption{Comparison of the Euclidean space fermion propagators in the Landau gauge and in the Yennie gauge. The Yennie gauge propagator is obtained directly from the $\alpha=1$ numerical solution for the spectral functions with $r_0=1$. The Landau gauge propagator is then calculated using the LKFT based on the Yennie gauge spectral functions. The top plot illustrates the propagator functions in the Euclidean space. The blue dash line is the Dirac vector part of the fermion propagator in the Landau gauge. The red dot line is the Dirac scalar part of the fermion propagator in the Landau gauge. The yellow solid line is the Dirac vector part of the fermion propagator in the Yennie gauge. The purple dot-dash line is the Dirac scaler part of the fermion propagator in the Yennie gauge. In the middle plot, the wavefunction renormalization of the fermion propagator is illustrated. The blue dash line is the wavefunction renormalization in the Landau gauge, while the red solid line is in the Yennie gauge. The bottom plot illustrates the mass function. The blue dash line represents the mass function in the Landau gauge, while the red solid line stands for the mass function in the Yennie gauge.}
\label{fig:LKFT_n10}
\end{figure}

As a parameterization of the numerical solutions, we explore the Pad\'{e} approximation of the spectral functions $r_j(s)$ after factoring out the asymptotics. In the parameterization we additionally take into the account the vanishing of the $r_j(s)$  in the $s\rightarrow m^2$ limit. After defining ${z=1-1/y}$, we propose the following parameterization of the spectral functions:
\begin{equation}
r_j(s)\simeq c_0\,y^{c_1}z^{c_2}\dfrac{1+\sum_{n=1} a_n z^n}{1+\sum_{n'=1} b_{n'} z^{n'}}.\label{eq:r_j_Pade}
\end{equation}
Recall that the asymptotic behavior of the numerical solutions $r_j(s)$ is given by Eqs.~\eqref{eq:asym_r1} and~\eqref{eq:asym_r2}, which fixes $c_1=-\gamma_1$. We then truncate the series expansions to ${n_{\mathrm{max}}=n'_{\mathrm{max}}=4}$, finding out an optimal set of parameters listed in Tab.~\ref{tab:Pade}. Notice from Fig.~\ref{fig:r1r2_n10} that the differences are typically less than a few percent.
\begin{table}[h]
	\caption{Fitting parameters for the Pad\'{e} approximations of the spectral functions $r_j(s)$ in the form of Eq.~\eqref{eq:r_j_Pade} when $\alpha=1$ and $r_0=1$.}\label{tab:Pade}
	\begin{ruledtabular}
		\begin{tabular}{c|cc}
			$\alpha=1,\,r_0=1$ & $\quad\quad r_1(s)\quad\quad $ & $\quad\quad r_2(s)\quad\quad $ \\ 
			\hline
			$c_0$ & $0.5017$ & $0.01969$ \\
			$c_1$ & $-0.7253$ & $-0.7253$ \\ 
			$c_2$ & $1.022$ & $1.993$ \\ 
			$a_1$ & $15070$ & $467.6$ \\ 
			$a_2$ & $-13540$ & $-25140$ \\ 
			$a_3$ & $175.0$ & $-5740$ \\ 
			$a_4$ & $85.26$ & $20630$ \\ 
			$b_1$ & $19690$ & $474.2$ \\ 
			$b_2$ & $-8469$ & $-26070$ \\ 
			$b_3$ & $-7689$ & $24340$ \\ 
			$b_4$ & $377.8$ & $-414.2$ \\
		\end{tabular}
	\end{ruledtabular}
\end{table}
\section{The Euclidean space fermion propagator in the Landau gauge from LKFT\label{ss:Landau}}
The LKFT provides the exact relation for the QED fermion propagator in different covariant gauges. The LKFT for the spectral representation of the fermion propagator is given by Eq.~(61) of Ref.~\cite{Jia:2016wyu}. We would like to apply the LKFT to calculate the Euclidean propagator functions in the Landau gauge using the the numerical spectral functions in the Yennie gauge we now have. 

Specifically, after removing divergences in the wavefunction renormalization, the LKFT for the fermion propagator becomes
\begin{align}
& \quad S_F(p^2,\xi)=\int ds\,\mathcal{K}_j(\xi-\xi')\dfrac{\rho_j(s;\xi-\xi')}{p^2-s+i\varepsilon}\nonumber\\
& =-\dfrac{\Gamma(n-\nu)}{\Gamma(n+\nu-1)}\int_{m^2}^{+\infty}ds\int_0^1 dt\,\dfrac{t^{-\nu} (1-t)^{n+\nu-2}}{(s+t\,p_{\mathrm{E}}^2)^{n-\nu}}\nonumber\\
&\quad \times\rho_j(s;\xi')s^{n-1},\label{eq:LKFT_S_F}
\end{align}
where $\nu=\alpha(\xi-\xi')/(4\pi)$. In the case of transforming the Yennie gauge spectral representations into the Landau gauge propagator functions, we have $\xi=0$ and $\xi'=3$.

In the numerical evaluation of Eq.~\eqref{eq:LKFT_S_F}, we need to apply the following integral representation of the hypergeometric functions to Eq.~(61) of Ref.~\cite{Jia:2016wyu} in order to regularize singularities in the $r_j(s)$ integrals:
\begin{align}
&\quad \,_2F_1(n-\nu;1-\nu;n;p^2/s)\nonumber\\
&=\dfrac{\Gamma(n)}{\Gamma(1-\nu)\Gamma(n-1-\nu)}\int_{0}^{1}dt\dfrac{t^{-\nu}(1-t)^{n+\nu-2}}{(1-tp^2/s)^{n-\mu}},
\end{align}
which is obtained from Eq.~(15.3.1) of Ref.~\cite{abramowitz1965handbook}. The resulting fermion propagators in the Landau gauge are given by Fig.~\ref{fig:LKFT_n10}. The divergent part of the wavefunction renormalization $Z_2=\exp[-\alpha\xi (1/\epsilon-\gamma_{\mathrm{E}}+\ln 4\pi)/(4\pi)]$ has been factored out in Fig.~\ref{fig:LKFT_n10}. In the Landau gauge, the finite part of the wavefunction renormalization is fixed such that $F(p^2)$ coincides with that in the Yennie gauge at $p^2=-10^{4}\,m^2$.

\section{Summary\label{ss:summary}}
Using a spectral representation for the fermion propagator, we have revised the solution of its SDE within a truncation where the fermion-boson vertex is given by a modification of the Gauge Technique of Salam, Strathdee, and Delbourgo~\cite{Delbourgo:1977jc,PhysRev.130.1287,PhysRev.135.B1398,PhysRev.135.B1428} to ensure loop-renormalizable in four dimensions. We find a series solution supplemented by the numerical coefficients in the Yennie gauge valid in the whole complex momentum plane for the resulting equations. This erratum corrects and supersedes the earlier results published in Ref.~\cite{Jia:2017niz}. Because the model is not gauge covariant, the on-shell renormalization conditions can not be satisfied away from the Yennie gauge. Consequently we applied the LKFT to obtained the Euclidean space fermion propagator in the Landau gauge based on the Yennie-gauge solution.
\section*{Acknowledgments}
This material is based upon work supported by the U.S. Department of Energy, Office of Science, Office of Nuclear Physics under contract DE-AC05-06OR23177 that funds Jefferson Lab research and under Grand No. DE-SC0023692. S.~J. is also supported by the U.S. Department of Energy, Office of Science, Office of Nuclear Physics, under Contract No. DE-AC02-06CH11357.
\appendix
\section{Simplifications of the fermion propagator SDE\label{ss:simplification_SDE_fermion}}
With an adjustable free-particle pole residue, Eq.~(A.1) of Ref.~\cite{Jia:2017niz} becomes
\begin{subequations}
\begin{align}
	& \lim\limits_{p^2\rightarrow m^2}\, [ S_2(p^2) + \overline{\sigma}_2(p^2) ] / S_1(p^2) = [ 1 - \lambda_2\alpha / ( 4\pi r_0 ) ] m, \\
	& \lim\limits_{ p^2\rightarrow m^2}\, \Big\{ p^2S_1(p^2) + \overline{\sigma}_1(p^2) - \dfrac{S_2(p^2)}{S_1(p^2)}[S_2(p^2) + \overline{\sigma}_2(p^2)] \Big\} \nonumber\\
	& = (\lambda_2-\lambda_1) \dfrac{\alpha}{4\pi} \lim\limits_{p^2\rightarrow m^2}\, \dfrac{m^2}{p^2-m^2} + q_1(m^2)-m\,q_2(m^2) \nonumber\\
	& + r_0 \! + \! [ 2 - \lambda_2\alpha / ( 4\pi r_0 ) ] [ m^2P_1(m^2) - mP_2(m^2) ].
\end{align}
\end{subequations}
Subsequently we have
\begin{subequations}
\begin{align}
& p^2 S_1(p^2) + \overline{\sigma}_1(p^2) = [ 1 - \lambda_2\alpha / ( 4\pi r_0) ] mS_2(p^2) \nonumber\\
& + (\lambda_2-\lambda_1) \dfrac{\alpha}{4\pi} \lim\limits_{p^2\rightarrow m^2}  \dfrac{m^2}{p^2-m^2} + q_1(m^2)-m\,q_2(m^2) \nonumber\\
& +r_0 + [ 2 - \lambda_2\alpha / ( 4\pi r_0 ) ] [m^2P_1(m^2)-mP_2(m^2)]\label{eq:SDE_RZ1_OnShell}\\
& S_2(p^2)+\overline{\sigma}_2(p^2) =  [ 1 - \lambda_2\alpha / ( 4\pi r_0 ) ] mS_1(p^2),\label{eq:SDE_RZm_OnShell}
\end{align}
\end{subequations}
with $q_j(p^2)$ defined in Eq.~\eqref{eq:def_bbSj}. The spectral function of $q_2(p^2)$ is given by
\begin{align}
	& \overline{R}_2(s) =-\dfrac{1}{\pi}\mathrm{Im}\{ q_2(s+i\varepsilon) \} = \dfrac{3\alpha r_0 m}{4\pi s} \theta(s-m^2) \nonumber\\
	& - \dfrac{\alpha\xi r_0 m^3}{4\pi s^2} \theta(s-m^2) - \dfrac{3\alpha}{4\pi}\left[\dfrac{4}{3}r_2(s) - \dfrac{1}{s} \int_{m^2}^{s}ds'\, r_2(s') \right] \nonumber\\
	& - \dfrac{\alpha\xi}{4\pi s^2} \int_{m^2}^{s}ds' \, s' r_2(s').\label{eq:def_R2_bar}
\end{align}
While for $\tilde{q}_1(q^2)$, based on Eq.~\eqref{eq:tilde_q1} we obtain 
\begin{align}
	& \tilde{\overline{R}}(s) = - \dfrac{1}{\pi}\mathrm{Im}\, \{ \tilde{q}_1(s+i\varepsilon) \} = \dfrac{3\alpha r_0 m^2 }{4\pi s^2} \theta(s-m^2) \nonumber\\
	& - \dfrac{\alpha\xi r_0}{4\pi s} \theta(s-m^2) - \dfrac{3\alpha}{4\pi}\left[ \dfrac{4}{3}r_1(s) - \dfrac{1}{s^2} \int_{m^2}^{s} ds'\, s' r_1(s') \right]\nonumber\\
	& - \dfrac{\alpha\xi}{4\pi s} \int_{m^2}^{s}ds'\,r_1(s').
\end{align}
\section{The subtracted spectral representation for the Dirac scalar component of the fermion propagator SDE\label{ss:sub_Diarc_scalar}}
\begin{figure}
\centering
\includegraphics[width=0.85\linewidth]{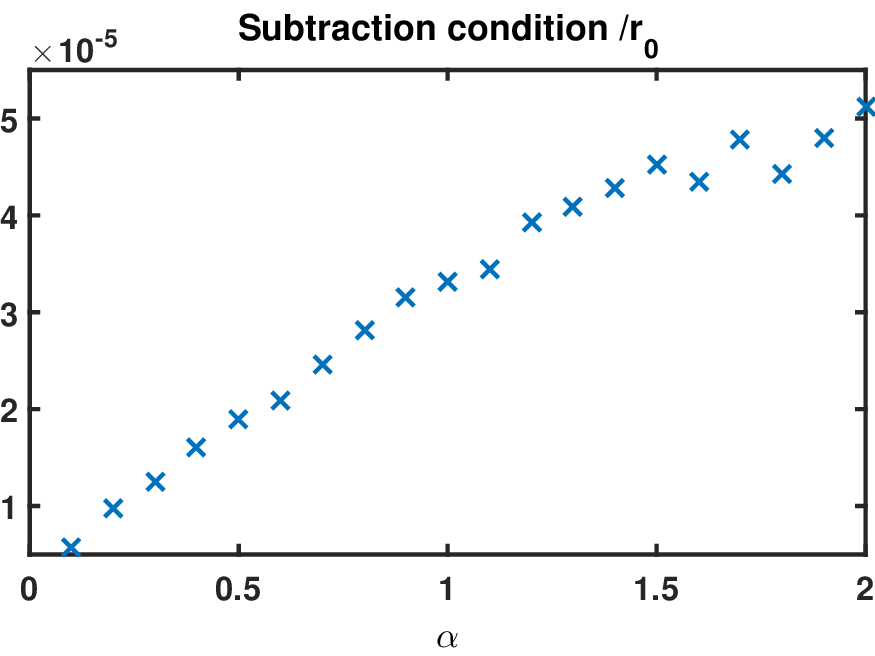}
\caption{Test of the numerical solutions by Eq.~\eqref{eq:SDE_r1_OnShell_p20_rdd} with different $\alpha$. Data points correspond to the absolute value of Eq.~\eqref{eq:SDE_r1_OnShell_p20_rdd} divided by $r_0$.}
\label{fig:Subt_Cond}
\end{figure}
Consider the spectral representation of a function depending on $p^2$. When the spectral function $\rho(s)$ is non-vanishing asymptotically, the spectral integral ${f(p^2)=\int ds\,\rho(s)/(p^2-s+i\varepsilon)}$ is not convergent. However, once we know the value of the function $f(p^2)$ at one point, for example $f(0)$, we can write down the subtracted spectral representation:
\begin{equation}
\tilde{f}(p^2) \equiv \dfrac{f(p^2)-f(0)}{p^2} = \int ds\dfrac{\rho(s)}{(p^2-s+i\varepsilon)\,s}.\label{eq:def_sr_subtracted}
\end{equation}
This subtracted spectral representation is then convergent as long as $\lim\limits_{s\rightarrow+\infty}\rho(s)/s\rightarrow 0$.

In order to ensure convergences of the spectral integrals in the $q_1$ terms of Eq.~\eqref{eq:SDE_r1_OnShell}, one subtraction is needed. We choose the subtraction point to be $p^2=0$. Such a subtraction also separates the renormalization constants embedded in Eq.~\eqref{eq:SDE_r1_OnShell} from the $p^2$ dependent terms. Specifically when $p^2=0$, Eq.~\eqref{eq:SDE_r1_OnShell} becomes
\begin{align}
& q_1(0) = ( 1 - \alpha / \pi ) mP_2(0) + ( 2 - \alpha / \pi ) \nonumber\\
& \times \left[m^2P_1(m^2)-mP_2(m^2) \right] + q_1(m^2)-mq_2(m^2).\label{eq:SDE_r1_OnShell_p20}
\end{align}
Subtracting Eq.~\eqref{eq:SDE_r1_OnShell_p20} from Eq.~\eqref{eq:SDE_r1_OnShell} and then multiplying $p^{-2}$ produce 
\begin{equation}
P_1(p^2) + \tilde{q}_1(p^2) = ( 1 - \alpha / \pi ) m \tilde{P}_2(p^2).\label{eq:Subt_Cond}
\end{equation}
From Eq.~\eqref{eq:def_sr_subtracted} we immediately have
\begin{equation}
\tilde{P}_2(p^2)=\int_{m^2}^{+\infty}ds\dfrac{r_2(s)}{(p^2-s+i\varepsilon)\,s}.
\end{equation}
While for $\tilde{q}_1(p^2)$, we apply the definition of $\tilde{q}_1(p^2)$ by Eq.~\eqref{eq:def_sr_subtracted} in the momentum space and obtain the following expression:
\begin{align}
	& \tilde{q}_1(p^2) = \dfrac{q_1(p^2)-q_1(0)}{p^2} = \dfrac{3\alpha r_0}{4\pi p^2} \left[ 1 + \dfrac{m^2}{p^2} \ln(1-p^2/m^2) \right] \nonumber\\
	& - \dfrac{\alpha\xi r_0}{4\pi p^2} \ln(1-p^2/m^2) - \dfrac{3\alpha}{4\pi} \int_{m^2}^{+\infty}ds\, \bigg\{\dfrac{4}{3}\dfrac{1}{p^2-s+i\varepsilon} \nonumber\\
	& - \dfrac{1}{p^2} \left[ 1 + \dfrac{s}{p^2} \ln(1-p^2/s) \right] \bigg\}r_1(s) - \dfrac{\alpha\xi}{4\pi p^2} \int_{m^2}^{+\infty}ds \, r_1(s) \nonumber\\
	& \times  \ln(1-p^2/s).\label{eq:tilde_q1}
\end{align}

Specifically when ${p^2=m^2}$ we have ${q_1(m^2)}=q_1(0)+m^2\tilde{q}_1(m^2)$. Consequently Eq.~\eqref{eq:SDE_r1_OnShell_p20} is reduced into
\begin{align}
	& ( 1-\alpha/\pi ) P_2(0) + ( 2 - \alpha/\pi ) [ mP_1(m^2) - P_2(m^2) ] \nonumber\\
	& + m\, \tilde{q}_1(m^2)-q_2(m^2)=0.\label{eq:SDE_r1_OnShell_p20_rdd}
\end{align}
Notice that Eq.~\eqref{eq:SDE_r1_OnShell_p20} is automatically satisfied in the massless limit. When evaluating ${m\,\tilde{q}_1(m^2)-q_2(m^2)}$, the ${\ln(1-p^2/m^2)}$ singularities cancel. Explicitly we have
\begin{align}
	& m\, \tilde{q}_1(m^2)-q_2(m^2) = \alpha(\xi+3)r_0 / ( 4\pi m ) \nonumber\\
	& - \dfrac{\alpha}{\pi} \int_{m^2}^{+\infty} ds \dfrac{m \, r_1(s)-r_2(s)}{m^2-s} + \dfrac{3\alpha}{4\pi m} \int_{m^2}^{+\infty} ds \Big[ 1 + \dfrac{s}{m^2} \nonumber\\
	& \times \ln(1-m^2/s) \Big] r_1(s) - \dfrac{3\alpha}{4\pi m^2} \int_{m^2}^{+\infty}ds\, \ln(1-m^2/s) \nonumber\\
	& \times r_2(s) - \dfrac{\alpha\xi}{4\pi m} \int_{m^2}^{+\infty}ds\, r_1(s) \ln(1-m^2/s) \nonumber\\
	& + \dfrac{\alpha\xi}{4\pi m^2} \int_{m^2}^{+\infty}ds \left[ 1 + \dfrac{s}{m^2} \ln(1-m^2/s) \right] r_2(s).
\end{align}
The $r_j(s)$ independent terms in $m\,\tilde{q}_1(m^2)-q_2(m^2)$ are only finite if the $\delta$-functions in $\rho_j(s)$ have the matching coefficients as given by Eq.~\eqref{eq:rhoj_delta_theta} (both being $r_0$).

In evaluating the spectral integrals in $q_j(p^2)$ numerically, the following reparameterization is recommended to ensure convergences:
\begin{align}
& \dfrac{1}{p^2} \int ds \left[ 1 + ( s / p^2 - 1 ) \ln( 1 - p^2/s ) \right]r_j(s)\nonumber\\
& = \int ds\int_{0}^{1}dt\dfrac{t-1}{tp^2-s+i\varepsilon}r_j(s).
\end{align}
\bibliography{manuscript_2F1_bib}
\end{document}